# Unipolar optical transitions in nanoclusters of ellipsoidal geometry

G. Nikoghosyan [a,*], H.S. Nikoghosyan [b]

[a] Yerevan State University, 1 Alex Manoogian Str., Yerevan, 0025, Armenia
[b] Shirak State University, 4 Paruyr Sevak Str., Gyumri, 3126, Armenia
[*] nikoghosyan@ysu.am

Abstract

The quantum states of an ellipsoidal nanocluster of a heterophase system $InAs/GaAs$ are studied using an exact analytical approach, in contrast to the generally accepted theoretical model based on the adiabatic approximation. It is shown that the spectrum of a nanoobject is formed from local groups, consisting of discrete levels, separated by terahertz frequency intervals. A double random degeneration of certain spectrum states is revealed. Allowed mid-infrared (IR) optical transitions between different spectral states are analyzed. The role of dimensional parameters and features of the shape of a nanoobject in the characteristics of unipolar transitions is assessed. The absorption spectrum for the transition in the lower part of the substructure spectrum is calculated.

Keywords: spectrum, radiation, range, degeneracy, oscillator, interval.

1. Introduction

The development of efficient semiconductor radiation sources in the mid-IR range is of great practical importance in key areas of science and technology. The creation of lasers of this kind is associated with certain difficulties, for the solution of which they usually resort to the use of growth technologies of heterostructures, superlattices and multicomponent solid solutions with a variable band gap. For these purposes, unipolar quantum cascade lasers, as well as interband lasers based on lead chalcogenides, are more often used. At the same time, the promise of using structures with quantum dots (QDs) in radiation sources in the noted range is obvious. This is due to the well-known advantages of QDs over structures with quantum wells with two-dimensional electron gas [1], [2]. Epitaxial crystallization methods provide the opportunity for targeted profiling of the geometry and holding potential of QDs for specific practical purposes. This is the key to the formation of an optimal spectrum of states and selection rules for interlevel transitions in order to ensure the desired radiative process, where the sources are the structural components of the nanomedium. In particular, during the self-organized growth of epitaxial $InAs$ on the $GaAs$ substrate, disk-shaped islands with vertical dimensions of several nanometers and lateral dimensions of several tens of nanometers ($20-100 nm$) are formed, which can be attributed to the shape of a highly oblate ellipsoid of revolution [3-8]. Below, for the model of a highly oblate ellipsoidal (HOE) QD system $InAs/GaAs$, within the framework of the approach [7], allowing an exact solution with complete separation of variables, the features of the energy spectrum, selection rules and oscillator strengths for intraband radiative transitions between levels of the lower part of the $c-$zone spectrum are studied (Fig.1). As a detailed theoretical analysis shows, the energy spectrum of an HOE cluster consists of discrete families of levels separated by wide forbidden intervals. Each family consists of closely spaced dimensional quantization levels, the interlevel intervals of which correspond to the terahertz range. At the same time, the fact that some of the discrete states corresponding to such levels are degenerate has been established. This work analyzes the nature of unipolar transitions under the conditions of quantum limitation of the ellipsoidal geometry of HOE QDs. Transitions between certain levels of neighboring discrete families in the lower part of the spectrum, allowed by selection rules, correspond to the mid-IR range. Such transitions, due to the possibility of suppressing the role of non-radiative processes, can become the basis for the generation of mid-IR radiation. Indeed, lattice absorption near the frequencies of optical phonons, due to the strong spatial limitation and localization of lattice vibrational modes in an ellipsoidal nanocluster, can be significantly suppressed with an appropriate choice of elastic parameters of the heterophase system. So, the mechanism of unipolar transitions under consideration can be used as a possible embodiment of laser lasing in the mid-IR range, using an HOE QD array as a working medium, as well as in quantum-sized IR photodetectors. It should be noted that for structures of the specified geometry, the adiabatic approximation is usually used, within which the concept of "fast" and "slow" subsystems is acceptable, representing, respectively, the movement along the short dimension of the ellipsoidal cluster and in its cross section. And when averaging the states of the "slow" subsystem over the states of the "fast" subsystem, motion in the transverse plane for weakly excited states is reduced to the model of a plane oscillator. This formulation of the problem leads to an equidistant spectrum of levels and to its own system of allowed and forbidden transitions, in accordance with certain selection rules [5], which are radically different from the results and physical consequences derived below for the exact solution. It is also worth pointing out a number of alternative approaches [9-10], which emphasizes the relevance of studying objects of the chosen geometry.

2. **Quantum states of the HOE cluster**

For the HOE QD model, with an isotropic quadratic law of electron dispersion, a quasi-spherical system of ellipsoidal coordinates $x = nr\sin\theta\cos\varphi, y = nr\sin\theta\sin\varphi, z = r\cos\theta$ is introduced, where the surface of the ellipsoid of revolution $\frac{x^2+y^2}{a^2}+\frac{z^2}{c^2}=1, a=cn$ has the form of a sphere with radius $r = c$ and $0 \leq \varphi \leq 2\pi, 0 \leq \theta \leq \pi, 0 \leq r \leq c$ [7].

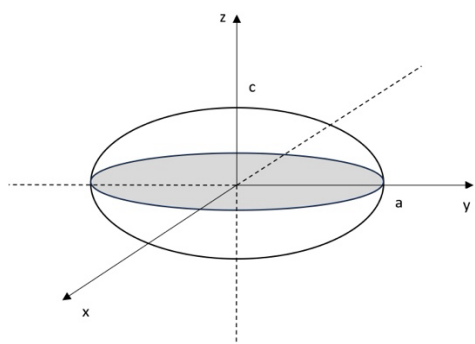

Fig.1. Ellipsoid of revolution diagram

Stationary states in a rectangular potential well of finite depth in the form of HOE QD in the effective mass approximation are determined by the Schrödinger equation.

$$-\frac{\hbar^2}{2m_c^*}\nabla^2\psi = E\psi$$

In the limiting case of a strongly oblate ellipsoid ($n \gg 1$, where $n = \frac{a}{c}$ is the degree of oblateness of the ellipsoid), by introducing a cylindrical radius $\rho = r\sin\theta, 0 \leq \rho \leq c$ instead of the polar angle $\theta$, we can proceed to orthogonal curvilinear coordinates $r, \rho, \varphi$, with a volume element $dV = n^2 \rho\, d\rho\, dr\, d\varphi$. The form of the Laplacian in the system $r, \rho, \varphi$

$$\nabla^2 = \frac{\partial^2}{\partial r^2} + \frac{1}{n^2}\left[\frac{1}{\rho}\frac{\partial}{\partial \rho}\left(\rho\frac{\partial}{\partial \rho}\right) + \frac{1}{\rho^2}\frac{\partial^2}{\partial \varphi^2}\right]$$

constructed using the Lamé coefficients $H_r = 1, H_\rho = n, H_\varphi = n\rho$ in the parabolic approximation, allows complete separation of variables and leads to a solution for the wave function of the particle

$$\psi = N\exp(iM\varphi)J_M(pn\rho)\cos(kr) \qquad (1)$$

Here $J_M$ – is the Bessel function. $k, p$ – are discrete components of the particle wave vector, which, based on the boundary conditions for the variables $r, \rho, \varphi$, determine the energy spectrum

$$E = E(k) + E(p) \qquad (2)$$

$N$ – is a normalization constant. That is, the spectrum of electrons in such QDs is formed from the levels of two possible states of motion in an ellipsoidal well. The first type of motion occurs along a short dimension of the disk, which corresponds to the basic structure of the spectrum of the electron $E(k) = \frac{\hbar^2 k^2}{2m_c^*}$, with large distances between levels. The conditions for the continuity of the wave function of the carrier motion along the radial direction of the sphere and its derivative on the surface of the sphere with radius $r = c = const$ in the system $r, \rho, \varphi$ lead to the standard dispersion equation

$$tg(kc) = \frac{q}{k}, q^2 = \frac{2m_{cB}^*}{\hbar^2}(\Delta E - E(k)) \qquad (3)$$

Here $m_{cB}^*$ – is the mass of the carrier in the matrix $GaAs$, $q$ – is the decay rate of the exponential decay of the wave function in $GaAs$. At $T = 0$, the magnitude of the potential step at the heterointerface in the $c-$zone $\Delta E_c = 0,7eV$, the effective mass of the electron in $InAs$ - $m_c^* = 0,027m_{0e}$ and in $GaAs$ - $m_{cB}^* = 0,065m_{0e}$, and at the average thickness of the cluster $2c = 24A°$, the solution of the dispersion equation (3) leads to the value of the energy $E(k) = 0,523eV$ at $kc = 0,73$. The second type of motion, not equivalent to the first, occurs in the transverse plane of the disk, forming a substructure of closely spaced levels of the spectrum of the QD $E(p) = \frac{\hbar^2 p^2}{2m^*}$, which is determined by the conditions of cyclicity of the wave function in the variables $\theta$ (i.e., in $\rho$) and in $\varphi$. When inverting spatial coordinates, the wave function (1) is transformed according to the $\psi \to (-1)^M \psi$ law. So in a one-particle problem, the parity of the state in $\rho$ coincides with the parity of the number $M$, where $M = 0,1,2,3,...$, due to the periodicity of $\psi$ in the variable $\varphi$. To demonstrate the cyclicity of the wave function $\psi$ in $\theta$ (i.e., in $\rho$), we choose the interval of change of $\theta$ as $-\frac{\pi}{2} \leq \theta \leq \frac{\pi}{2}$, and accordingly $-c \leq \rho \leq c$. Then the conditions for the cyclicity of the wave function in $\theta$ along the surface of the QD, in accordance with (1), are represented in the form

$$J_M(pnc) = J_M(-pnc) \quad (4)$$

$$J'_M(pnc) = J'_M(-pnc) \quad (5)$$

Since for odd $M$

$$J_M(-pnc) = -J_M(pnc)$$

then the cyclicity condition (4) leads to a dispersion equation for determining the substructure levels for odd $M$

$$J_M(\Upsilon_S(M)) = 0, \Upsilon_S(M) = pnc = pa, M = 1,3,... \quad (6)$$

Here $\Upsilon_S(M)$ – are the dimensionless roots of the Bessel functions, which are characterized by the number $S$. For even $M$ we have $J'_M(-pnc) = -J'_M(pnc)$, so condition (5) leads to the conclusion

$$J'_M(Z_S(M)) = 0, Z_S(M) = pnc = pa, M = 0,2,4,... \quad (7)$$

where $Z_S(M)$ – are the dimensionless roots of the derivative of the Bessel functions. That is, for even $M$, condition (7) can be used as a dispersion equation. As a result of the analysis in Table 1, Fig. 2 and Fig. 3 present a fragment of the lower part of the spectrum of the $c-$ zone of the HOE QD in the form of tabulated energy values and the relative location of the electronic levels $E_{(f,M,S)} = E(k) + E(p)$ of the substructure, corresponding to a certain (lowest) level of the main structure. $f-$ is a quantum number characterizing the level of the main structure, $M,S-$ are the quantum numbers of substructure levels. In this case, in a QD of composition $InAs$ in the matrix $GaAs$, only one electronic level of the main structure, i.e. $f = 1$, is usually placed.

|  |  | $E_{(115)} = 0,9615eV$ | $E_{(125)} = 0,9473eV$ | $E_{(135)} = 1,334eV$ |
|---|---|---|---|---|
| $E_{(105)} = 0,6289eV$ | $E_{(114)} = 0,6289eV$ | $E_{(124)} = 0,6157eV$ | $E_{(134)} = 0,9325eV$ |
| $E_{(104)} = 0,3669eV$ | $E_{(113)} = 0,3669eV$ | $E_{(123)} = 0,3516eV$ | $E_{(133)} = 0,6eV$ |
|  |  |  | $E_{(132)} = 0,3376eV$ |
|  |  |  |  |
| $E_{(103)} = 0,1745eV$ | $E_{(112)} = 0,1745eV$ | $E_{(122)} = 0,1591eV$ |  |
| $E_{(102)} = 0,052eV$ | $E_{(111)} = 0,052eV$ | $E_{(121)} = 0,033eV$ | $E_{(131)} = 0,1443eV$ |
| $M = 0$ | $M = 1$ | $M = 2$ | $M = 3$ |

Table 1. Energy levels of the $c-$ zone substructure with cluster parameters $2c = 24 A°$ $2\bar{a} = 400 A°$ ,corresponding to the size-quantized level of the main structure $f = 1$.

Note that the substructure levels that actually appear in the QD HOE are located in the $I(k) = \Delta E_c - E(k) = 0,177 eV$ energy interval (lower part of Table 1). So, given the thickness $2c = 24 A°$, diameter $D = 2\bar{a} = 400 A°$ and the value of the interval $I(k)$, a total of 5 electronic levels $E_{(f,M,S)}$ are placed in the cluster in the parabolic approximation for the energy spectrum of electrons in $InAs$: $E_{(121)}, E_{(111)}, E_{(131)}, E_{(122)}, E_{(112)}$. And of these levels

$$E_{(102)} = E_{(111)}, E_{(103)} = E_{(112)} \tag{8}$$

are doubly degenerate, which is obvious from the relationship between the zeros of the Bessel function and the derivative of the Bessel function $Z_S(0) = \Upsilon_{S-1}(1)$ [8]. The doubly degenerate levels of the HOE QD spectrum correspond to the following pairs of eigenfunctions

$$\{(102),(111)\};\{(103),(112)\} \tag{9}$$

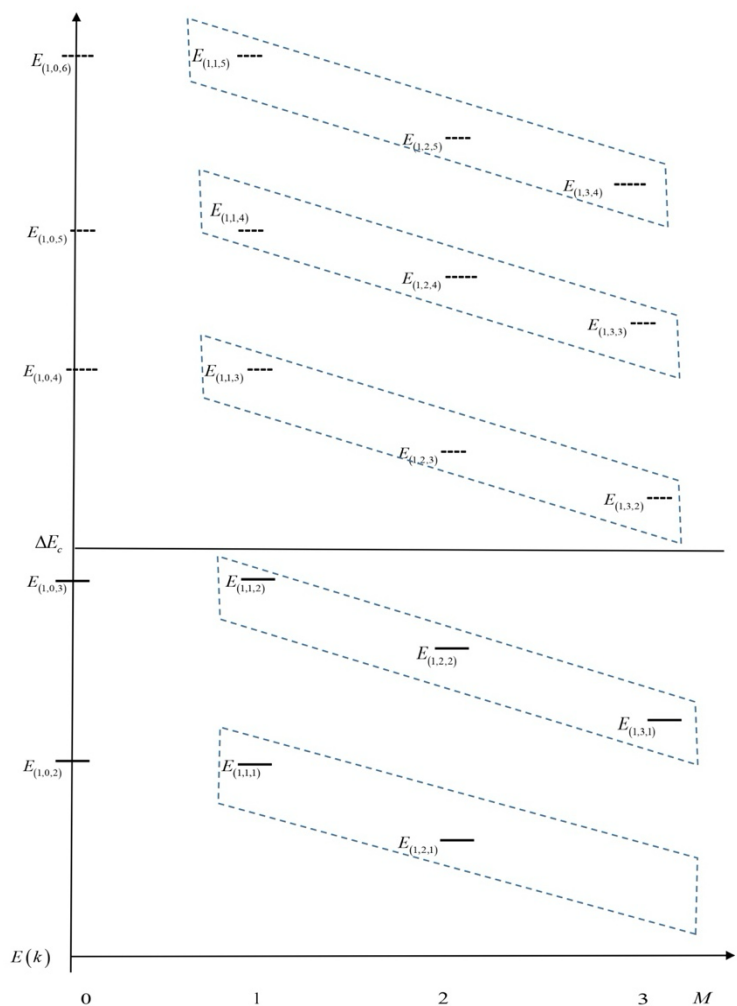

Fig.2. Configuration of the spectrum of the $c-$ zone of the HOE QD depending on the quantum numbers $f = 1, M, S$ for the cluster parameters $2c = 24 A°$ and $2\bar{a} = 400 A°$.

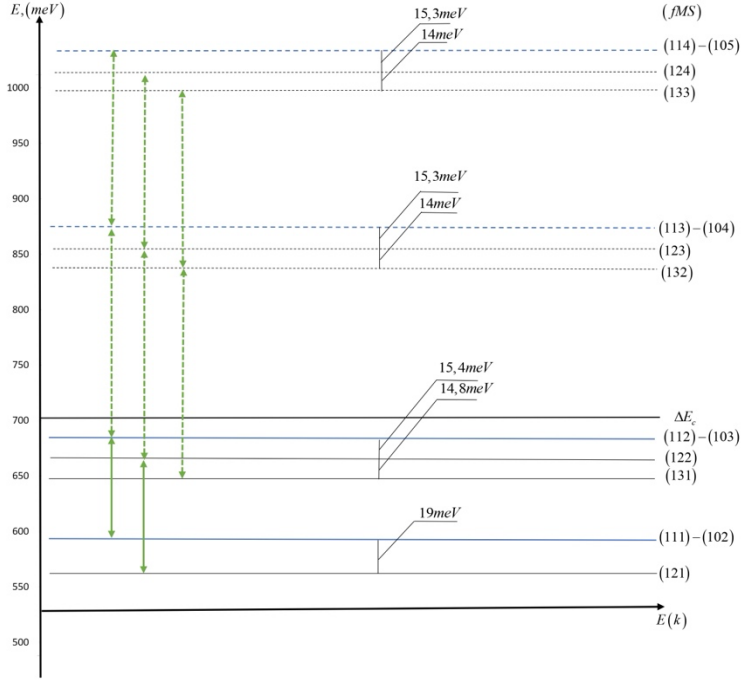

Fig.3. Energy level diagram and allowed transitions of the HOE cluster for parameters $2c = 24 A°$, $2\bar{a} = 400 A°$ and $f = 1$.

### 3. Electrodipole transitions between states of the $c$-zone of the size-quantized spectrum of the HOE QD

The nature of intraband dipole electrical transitions between states of the substructure is determined by the strength of the transition oscillator $m \to k$

$$F_{km}^{\rho} = \frac{2 m_c^* \omega_{km}}{\hbar} |\mu_{km}|^2 \qquad (10)$$

here $k = (fM'S')$ and $m = (fMS)$ denote the states of different levels of the $c$-zone $(k \neq m)$, the state functions of which are orthogonal at $M \neq M'$, where $f = f' = 1$, $\omega_{km}$ — are the transition frequency, and $\mu_{km} = \langle k|\rho|m\rangle$. Here, for non-degenerate states we have

$$\mu_{km} = 2\pi a^2 c N_1 N_2 I_0 \int_0^1 J_M^* (W_{S'}(M) x) J_M (W_S(M) x) x^2 dx,$$

where $W_S(M) = \Upsilon_S(M)$, for odd $M$, $W_S(M) = Z_S(M)$ for even $M$, $x = \frac{\rho}{c}$, $I_0 = \frac{c}{2}\left(1 + \frac{\sin 2kc}{2kc}\right), c = 1,2 nm$ — is the thickness of the HOE cluster, $N_{1,2} = \dfrac{1}{\left(2\pi a^2 \int_0^1 [J_M(W_S(M)x)]^2 x dx \cdot I_0\right)^{1/2}}$ — are normalization constants.

In the case of a transition between degenerate states

$$|\mu_{km}|^2 = \sum_{i=1}^{g_1} \sum_{j=1}^{g_2} |\mu_{ij}|^2, \tag{11}$$

where $g_1, g_2$ –are the multiplicities of degeneracy of the levels involved in the dipole transition. In the lower part of the spectrum of the HOE QD substructure, one can note a series of allowed optical transitions, the intensities of which can be characterized by the values of the oscillator strengths (Table 2).

| Wavelength, *mkm* | Transition | Oscillator strength $|F_{km}^\rho|$ |
|---|---|---|
| 2,9621 | $\{(102),(111)\} \to \{(103),(112)\}$ | 0,00332 |
| 2,8775 | $(121) \to (122)$ | 0,00255 |
|  |  |  |
|  |  |  |
| 1,8850 | $(122) \to (123)$ | 0,00548 |
| 1,8859 | $\{(103),(112)\} \to \{(104),(113)\}$ | 0,01527 |
| 1,8772 | $(131) \to (132)$ | 0,00339 |
| 1,3849 | $\{(104),(113)\} \to \{(105),(114)\}$ | 0,02126 |

Table 2. Oscillator strength for some transitions in the lower part of the spectrum of the HOE QD substructure.

In a nanocluster with parameters $2c = 24 A°$, $2\bar{a} = 400 A°$, $I(k) = 0,177 eV$, possible transitions in the mid-IR range occur (upper part of Table 2)

$$\{(102),(111)\} - 122,5 meV \to \{(103),(112)\}; (121) - 126,1 meV \to (122).$$

Moreover, for the transition between non-degenerate states $(121) \to (122)$ we have

$$|\mu_{km}|^2 = |\langle (122)|\rho|(121)\rangle|^2,$$

and for the transition between degenerate states $\{(102),(111)\} \to \{(103),(112)\}$

$$|\mu_{km}|^2 = |\langle (103)|\rho|(102)\rangle|^2 + |\langle (112)|\rho|(111)\rangle|^2$$

It is obvious that by varying the structural parameters $a, c, I(k)$, it is possible, for practical purposes, to increase the number of states of the substructure corresponding to a certain size-quantized level of the main structure, which actually appear in the quantum well of the HOE QD. This will make it possible to use a system of discrete families of substructure levels to study, in particular, cascade processes of carrier relaxation under conditions of size quantization.

Of great practical importance is the identification of possible changes in the charge configuration of the system during unipolar transitions between states of individual families of levels. This purpose can be served, first of all, by calculating the dipole moments of the states of HOE QDs in the transverse direction, perpendicular to the growth axis of the structure. Since solution (1) has a certain parity, it is obvious that when inverting coordinates, the probability density does not change. That is, the charge distribution in non-degenerate states (1) has a center of symmetry and the electric dipole moment in the transverse plane is equal to zero. A similar picture occurs for degenerate states of HOE QDs. This can be verified by constructing for degenerate quantum states (9) orthogonal superposition basis vectors without a certain parity, such as the combination $\psi_{i(j)} = \dfrac{\psi_{(102)} \pm \psi_{(111)}}{\sqrt{2}}$, and calculating the dipole moments in the transverse direction $d = e \int \psi^*_{i(j)}(r,\rho,\varphi) \rho \psi_{i(j)}(r,\rho,\varphi) dV$. Here, in contrast to the hydrogen atom, which is characterized by $l-$ degeneracy ($l-$ is the orbital quantum number), transitions between the states of the degenerate pair $\psi_{(102)}$ and $\psi_{111}$ are prohibited by selection rules. So there is no mixing of degenerate states with the same energy $\psi_{(102)}$, $\psi_{111}$ and, accordingly, the formation of their own dipole moment [11-14]. In the group of states belonging to degenerate levels (8), the electric moment can appear only as a result of polarization in an external electric field, leading to splitting of the levels and removal of degeneracy.

During electric-dipole transitions between levels of the substructure, in an axially symmetric field of an ellipsoidal cluster, the law of conservation of the projection of the total moment is satisfied

$$J'_z = J_z + J_{zph}, \qquad (12)$$

where $J_z, J'_z -$ are the projections of the total moment on the $z$ axis, respectively, for the initial and final states of the quantum system involved in the transition, $J_{zph} -$ is the projection of the total moment of the electric dipole radiation. Taking into account the fact that the spin state of a quantum system does not change during a dipole electric transition, (12) takes the form (in units of $\hbar$)

$$M' = M + J_{zph}, \qquad (13)$$

Equation (13) is consistent with the selection rule for the quantum number $M$ only in the case of radiation linearly polarized in the transverse plane, for which $J_{zph} = 0$.

4. **Optical absorption in an HOE QD array**

From the point of view of the practical implementation of "self-organized" structures during epitaxy of strongly mismatched materials $InAs$ and $GaAs$, the analysis of optical absorption during unipolar transitions in HOE QDs is relevant. This is important, in particular, for the design of such unipolar device structures as quantum-sized IR photodetectors. It is also of interest to study the influence of the anisotropy factor of the quantum confinement shape on the absorption coefficient and sensitivity of photoreception in the IR range during transitions between states of the substructure of the HOE QD spectrum. Let's calculate the absorption coefficients using the following expression [15] (Fig. 4)

$$\alpha = \dfrac{(N_j - N_i)|\vec{d}_{ij}|^2 \omega_{ij}}{2\hbar c_0 n_0 \varepsilon_0} g_G(\omega, \omega_{ij}) \qquad (14)$$

where $N_i, N_j$ – are the concentrations of QDs in the states involved in the transition $|j\rangle \to |i\rangle$, $g_G(\omega, \omega_{ij})$ – is the spectral (Gaussian) form factor describing the inhomogeneous broadening of the absorption spectrum of QDs due to their size dispersion, $n_0$ – is the refractive index of the medium, $c_0$ – is the speed of light, $\varepsilon_0$ – is the electrical constant, $\Delta\omega_G$ – is the total width Gaussian line at half-maximum level. The value of the absorption coefficient at the center of the broadening band is obtained by substituting $g_G^{max} = \dfrac{4\sqrt{\pi \ln 2}}{\Delta\omega_G}$ into (14). And the maximum absorption during an allowed transition from the lowest state $(121) \to (122)$ is the following

$$\alpha_{max} = 2\sqrt{\pi \ln 2}\, \frac{N|\vec{d}_{ij}|^2 \omega_{ij}}{\hbar \Delta\omega_G n_0 c_0 \varepsilon_0} \tag{15}$$

where $j = (121), i = (122)$, $|\vec{d}_{ij}| \sim e \cdot 0{,}169\,nm$, $\omega_{ij} = 1{,}9158 \cdot 10^{14}\,s^{-1}$, $n_0 = 3{,}3$, $N$ – is the concentration of absorbing centers (near equilibrium $N = N_{(121)} \gg N_{(122)}$). Table (3) shows the values of the absorption coefficient in the center of the spectral absorption band at the volume concentration of QDs $10^{17}\,cm^{-3}$ when $\Delta\omega_G \sim (0{,}15\cdot 10^{14} - 1{,}5\cdot 10^{14})\,s^{-1}$.

| $\Delta\omega_G\,(s^{-1})$ | $\alpha_{max}\,(cm^{-1})$ |
|---|---|
| $0{,}15\cdot 10^{14}$ | 29,76 |
| $0{,}675\cdot 10^{14}$ | 6,847 |
| $1{,}5\cdot 10^{14}$ | 2,98 |

Table 3. Calculated values of $\alpha_{max}$ in an environment with HOE QD.

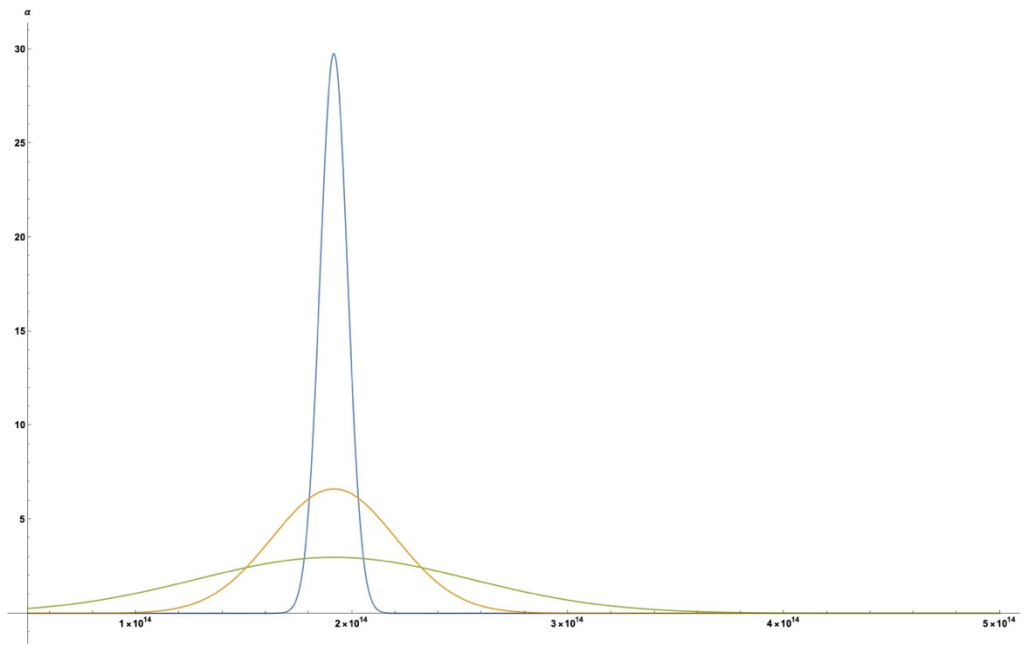

Fig.4. Absorption line for the $(121) \rightarrow (122)$ transition in the HOE QD medium of the heterophase system $InAs/GaAs$ :blue line - $\Delta\omega_G = 0,15 \cdot 10^{14} s^{-1}$, yellow line - $\Delta\omega_G = 0,675 \cdot 10^{14} s^{-1}$, green line - $\Delta\omega_G = 1,5 \cdot 10^{14} s^{-1}$.

## 5 The discussion of the results

Analysis of quantum states shows that the spectrum of the HOE QD substructure, related to a certain $(f=1)$ level of the main structure, is formed from discrete families of $\underbrace{E_{121}, E_{111}}_{\varepsilon_1}; \underbrace{E_{131}, E_{122}, E_{112}}_{\varepsilon_2}; \underbrace{E_{132}, E_{123}, E_{113}}_{\varepsilon_3}; \ldots$ levels, separated by wide forbidden intervals $\varepsilon_2 - \varepsilon_1 = 92,3 meV; \varepsilon_3 - \varepsilon_2 = 163,1 meV; \ldots$. Moreover, certain levels of each family $(E_{111}, E_{112}, E_{113}, E_{114}, E_{115}, \ldots)$ are doubly degenerate. In particular, the spectrum of the considered nanoobject of the heterophase system $InAs/GaAs$ contains the first two families $\varepsilon_1$ and $\varepsilon_2$. Each family consists of closely spaced dimensional quantization levels, the interlevel intervals of which correspond to the terahertz range. It can be assumed that with increasing state energy, the localization region of the electron narrows, moving away from the transverse plane of the maximum cross section of the HOE QD. This in turn leads to an increase in the degree of overlap of the wave functions of the states involved in unipolar transitions, as a consequence of the convergence of their localization regions and, accordingly, to an increase in the strengths of the transition oscillator as one moves up the energy scale (Table 2). The fact of random degeneration of the levels noted above determines a number of features of the system's behavior. Thus, a comparison of the data in Table 2 leads to the conclusion that transitions between neighboring degenerate
$\{(102),(111)\} \rightarrow \{(103),(112)\}; \{(103),(112)\} \rightarrow \{(104),(113)\};$
$\{(104),(113)\} \rightarrow \{(105),(114)\}; \ldots$
levels are more likely than transitions between non-degenerate levels of the corresponding families. In this case, the strength of the transition oscillator between degenerate levels increases with increasing family numbers. The discussed unipolar optical transitions can only occur with the participation of radiation linearly polarized in the transverse plane, as a consequence of the laws of parity conservation and the projection of the total moment onto the symmetry axis of the system. As a result of the calculations carried out, one can also note the revealed analytical dependence of the characteristics (oscillator strength) of unipolar transitions on the dimensional parameters and geometry of the HOE QD. In particular, in addition to the implicit one, there is also an explicit inversely proportional dependence of the strength of the transition oscillator on the degree of flattening $n$ of the HOE QD within the feasibility of condition $n \gg 1$. This makes it possible to control the characteristics of transitions by varying the dimensional parameters $a$ and $c$. The ellipsoidal geometry and features of quantum confinement, which shape the dynamics of the behavior of carriers in the nanoclusters under consideration, can lead to other physical consequences that are of practical interest. In particular, the characteristic frequencies of motion along the short dimension of the ellipsoid of rotation (the "fast" subsystem) are greater than the characteristic frequencies of motion in the transverse plane (the "slow" subsystem). Accordingly, the frequency of "impacts" of the carrier on the barriers and the probability of tunneling from the potential well of the HOE QD in the longitudinal (along the axis of structure growth) direction are higher than in the transverse direction. This difference becomes more significant in the case of applying an electric field in the longitudinal direction. Carriers filling the working levels of the HOE QD by illuminating radiation with a frequency in the interband absorption band appear in a state lying closer to the edge of the continuum. Here, the width of the barrier separating discrete states and states of a continuous spectrum is smaller than in the absence of an electric field. This can lead to an effective increase in tunneling penetration through the potential barrier (QD ionization) and an increase in the photocurrent. Of course, the tunneling coherence requirement must be met. The tunneling lifetime of an electron at the resonance level $\tau \sim \hbar/\Gamma$, where $\Gamma-$ is the width of the resonance level, must be less than the characteristic relaxation time of the electron along the momentum $\tau_p$, taking into account all scattering mechanisms. The allowed optical transition $(121) \to (122)$, which we considered in Section 4, occurs between the levels of different discrete groups of the spectrum of the HOE QD substructure. To compare the results obtained with known data, we present the value of the dipole moment of transition between neighboring levels of a harmonic oscillator for the model of a parabolic potential well in a spherical QD, with an energy difference $\hbar\omega_{ij} = 0,13 eV$, calculated in [2], $|\vec{d}| \sim e \cdot 1,15 nm$. And the calculation for a two-dimensional infinitely deep well of width $L$ for the transition between the ground and first excited levels gives $|\vec{d}_{10}| \sim e \cdot 0,18 L$ [16]. So the underestimated value of the dipole moment (and, accordingly, the absorption coefficient) that we obtained is apparently explained by the peculiarities of the overlap of the wave functions of the states involved in the optical transition, which have different functional characteristics of spatial localization. Note that the absorption coefficient exhibits (like the strength of the transition oscillator), in addition to an implicit one, also an explicit inversely proportional dependence on the degree of oblateness $n$ of the HOE QD within the limits of the feasibility of condition $n \gg 1$. So, in order to enhance absorption, you can vary the dimensional parameters of the structural units of $a, c$. Of course, it is also necessary to reduce the spectrum width due to a narrower size distribution of QDs. It is obvious that more efficient absorption is observed in optical transitions between degenerate states (type $\{(102),(111)\} \to \{(103),(112)\}$), provided that the lower operating level of the transition is filled by illuminating radiation with a frequency in the interband absorption band. An important factor in structural analysis is the effects of deformation under elastic stresses in a heterophase system $InAs/GaAs$. It should be noted that due to the difference in the lattices (with a mismatch $7\%$ for a pair $InAs/GaAs$), the overgrown islands $InAs$ in the matrix $GaAs$ are coherent elastic inclusions that create long-range elastic stress fields throughout the overgrown heterophase system, when the distance between the islands becomes comparable to the sizes of the islands [1]. The deposited matrix material above the islands is elastically stressed and this can affect the band characteristics of the materials. The anisotropy of the strain distribution in the heterophase system $InAs/GaAs$ leads to a displacement of the edges of the conduction band of the heteropair in the center of the Brillouin zone. Uneven shifts of the edges of the $c-$band at the heterointerface lead to a decrease in the depth of the potential well and the binding energy of the electron in $InAs$. In addition to the inhomogeneous spatial distribution of elastic deformations, factors such as the anisotropic piezoelectric potential associated with the appearance of polarization charges due to shear deformations at the $InAs/GaAs$ interfaces, also influence. However, the wave functions of the lowest states of the electron are localized at distances $|r| \ll c$, i.e., far from the ellipsoidal surfaces of the QD. So the effects at the interfaces of the heteropair $InAs/GaAs$ have little effect on the characteristics of the unipolar transitions considered above. This conclusion is also supported by the relatively large depth of the potential well $\Delta E_c = 0,7 eV$.

## 6  Conclusion

The geometric features of quantum limitation in HOE QD predetermine the use of the well-known analytical approach of complete separation of variables by introducing an orthogonal quasi-spherical coordinate system. This leads to a classification of the forms and characters of movement states, which forms the structure of the spectrum, consisting of separate groups of discrete levels. Allowed unipolar transitions between degenerate levels of neighboring lowest groups, corresponding to the mid-IR range, are characterized by relatively large values of the transition oscillator strengths and occur with the participation of radiation linearly polarized in the transverse plane. The charge distribution in the quantum states of HOE QDs, while not spherically symmetric, has a center of symmetry, which is preserved during unipolar transitions. The anisotropy of the quantum confinement shape is reflected in the absorption coefficient during unipolar transitions, revealing the structural dependence of $\sim \dfrac{1}{n^2}$.